\documentclass[traditabstract]{aa}

\usepackage{epsfig}
\usepackage{lscape}
\usepackage{natbib}
\bibpunct{(}{)}{;}{a}{}{,}    
\usepackage{graphics}
\usepackage{graphicx,graphics}
\usepackage{color} 
\usepackage{ulem}
\usepackage{gensymb}
\usepackage{times}
\usepackage{amsmath}
\usepackage{amssymb}
\begin{document}

\title{Peculiar motions of the gas at the centre of the barred galaxy UGC~4056}

\author{
        I.~A.~Zinchenko\inst{\ref{MAO},\ref{ARI}} \and
        L.~S.~Pilyugin\inst{\ref{MAO},\ref{ARI}} \and
        F.~Sakhibov\inst{\ref{UASM}} \and
        E.~K.~Grebel\inst{\ref{ARI}} \and
        A.~Just\inst{\ref{ARI}} \and
        P.~Berczik\inst{\ref{JU},\ref{ARI},\ref{MAO}} \and
        Y.~A.~Nefedyev\inst{\ref{KFU}} \and
        J.~M.~V\'{i}lchez\inst{\ref{IAA}} \and
        V.~M.~Shulga\inst{\ref{JU},\ref{IRA},\ref{CJU}}
       }
       
\institute{
Main Astronomical Observatory, National Academy of Sciences of Ukraine, 
27 Akademika Zabolotnoho St., 03680, Kyiv, Ukraine\label{MAO}
\and
Astronomisches Rechen-Institut, Zentrum f\"{u}r Astronomie der Universit\"{a}t Heidelberg, 
M\"{o}nchhofstr. 12-14, 69120 Heidelberg, Germany \label{ARI} 
\and
University of Applied Sciences of Mittelhessen, Campus Friedberg,
Department of Mathematics, Natural Sciences and Data Processing,
Wilhelm-Leuschner-Stra\ss e 13, 61169 Friedberg, Germany \label{UASM} 
\and
Kazan Federal University, 18 Kremlyovskaya St., 420008, Kazan, Russian Federation \label{KFU}
\and
Instituto de Astrof\'{i}sica de Andaluc\'{i}a, CSIC, Apdo, 3004, 18080 Granada, Spain \label{IAA}
\and
The International Center of Future Science of the Jilin University, 2699 Qianjin St., 130012, Changchun City, China \label{JU}
\and
Institut of Radio Astronomy of National Academy of Sciences of Ukraine, 4 Mystetstv str., 61002 Kharkov, Ukraine \label{IRA} 
\and
College of Physics, The Jilin University, 2699 Qianjin St., 130012, Changchun, China \label{CJU}
}

\abstract{We derive the circular velocity curves of the gaseous and stellar discs of UGC~4056,
a giant barred 
galaxy with an active galactic nucleus (AGN). We analyse UGC~4056 using the 2D spectroscopy obtained within 
the framework of the Mapping Nearby Galaxies at APO (MaNGA) survey. Using images and the colour index $g-r$ from 
the Sloan Digital Sky Survey (SDSS), we 
determined the tilt of the galaxy, which allows us to conclude that the galaxy 
rotates clockwise with trailing spiral arms. We found that the gas motion at the central 
part of the UGC~4056 shows peculiar features. The rotation velocity of the gaseous disc 
shows a bump within around three kiloparsecs while the rotation velocity of the stellar 
disc falls smoothly to zero with decreasing galactocentric distance. 
We demonstrate that the peculiar radial velocities in the central part of the galaxy 
may be caused by the inflow of the gas towards the nucleus of the galaxy. 
The unusual motion of the gas takes place at the region with the AGN-like radiation
and can be explained by the gas response to the bar potential.}



\titlerunning{Peculiar motions of the gas in UGC~4056}
\authorrunning{Zinchenko et al.}
\maketitle

\section{Introduction}

\begin{figure}
\resizebox{1.00\hsize}{!}{\includegraphics[angle=000]{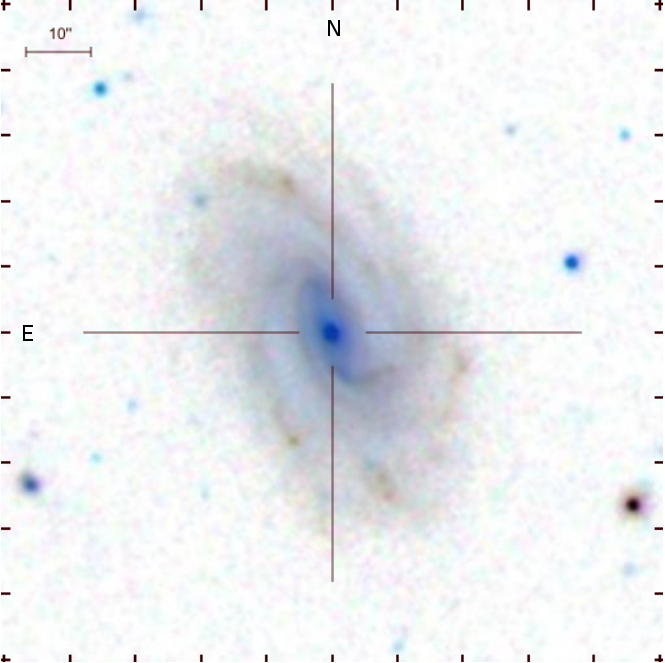}}
\caption{Image of the galaxy UGC 4056 in the SDSS $gri$ filters.}
\label{figure:image}
\end{figure}

Optical and radio observations of nearby galaxies show evidence of both gas inflow towards the 
active galactic nuclei
\citep[AGNs;][]{Knapen2000,StorchiBergmann2007,Riffel2008,Riffel2013,Combes2014,Davies2014,Lena2015} 
and gas outflow from the AGNs \citep{StorchiBergmann2007,Barbosa2009,Davies2014}
on the scales of $\sim 10-100$~pc. Gas inflow on these scales is frequently associated
with nuclear bars and spiral arms. 
However, the impact of the gas inflow and outflow on
the circum-nuclear region of galaxies and AGNs is still unclear. 

A peculiar feature in the velocity field of the gaseous disc was revealed in the central part of UGC~4056,
the only galaxy among more than 150 galaxies from the  Mapping Nearby Galaxies at APO (MaNGA) survey 
for which circular velocity curves were derived \citep{Pilyugin2018b}. This motivates us to 
consider this galaxy in detail and study the gas kinematics on scales of a few kiloparsecs around the AGN.

The galaxy UGC~4056 is a giant spiral galaxy of the morphological type SAB(s)c
\citep{RC3} and hosts an AGN.    
Two-dimensional spectroscopy of UGC~4056 was carried out within the MaNGA
survey project, which is a part of the Sloan Digital Sky Survey \citep{Albareti2016}. 
The distributions of the surface brightness, the emission line fluxes, 
and the wavelengths of the emission and absorption lines (and consequently 
of the line-of-sight velocity field) across the image of this galaxy can be obtained 
from the MaNGA 2D spectroscopy. The moderate value of the inclination 
angle of the galaxy UGC~4056 (Fig.~\ref{figure:image}) facilitates the 
determination of the photometric (the surface brightness profile, 
colour profile), chemical (the radial abundance gradient), 
and kinematic (the circular velocity curve) properties from the analysis 
of the MaNGA 2D spectroscopy.

The paper is structured as follows. The data are discussed in Section~\ref{sect:data}. 
In Section~\ref{sect:Morphology} we investigate the morphology of the detected peculiar 
motions using distributions of observed line-of-sight velocities of gas and stars 
along the major and minor axis of UGC~4056. The determination of the circular velocity curve, 
position angle (PA), and inclination of the target galaxy using a simple rotation 
model is presented in Section~\ref{sect:Rotation curve of M-8140-12703}. An 
application of the Fourier analysis of the azimuthal distribution of the non-circular 
motions in thin ring zones at different galactocentric distances in the plane 
of the galaxy is presented in Section~\ref{section:Fourier}. Our results are summarised 
in Summary.

\section{Data}
\label{sect:data}

The giant spiral galaxy UGC~4056 
is located at a redshift of $z$=0.03203, and the corresponding distance
is 133.3 Mpc (the NASA/IPAC Extragalactic Database,
{\sc ned})\footnote{The NASA/IPAC Extragalactic Database ({\sc ned}) is
operated by the Jet Propulsion Laboratory, California Institute of
Technology, under contract with the National Aeronautics and Space
Administration. {\tt http://ned.ipac.caltech.edu/}}. 
The adopted distance was obtained by NED after applying a correction for infall
towards the Virgo cluster, the Great Attractor, and the Shapley concentration 
with H$_0 = 73$ km/s/Mpc.
With this distance, the spatial scale of the MaNGA data corresponds to 624 pc/arcsec.
According to \citet{RC3}, the galaxy has the morphological type SAB(s)c.
\citet{Baillard2011} confirm the presence of a bar, with a length of about one third of the isophotal diameter.
Spiral arms and a weak bar are visible in Fig.~\ref{figure:image}. 
The bar extends up to a radius of $\sim$ 8 arcsec (which corresponds to 5~kpc in linear scale) from the centre of 
the galaxy roughly along the major axis of the disc. The spiral arms begin at the ends of the bar.
However, the catalogue of visual morphological classification by \citet{Nair2010} based on SDSS images
does not confirm the presence of a bar and assigned UGC~4056 the morphological type of T=4 with an inner lens.

The stellar mass of the galaxy UGC~4056 is $M_{s}$ = $10^{11.22}$M$_{\sun}$ 
according to the SDSS database (table {\sc stellarMassPCAWiscBC03}). 
This value was obtained via the principal component analysis (PCA) method \citep{Chen2012} 
using the library of model spectra by \citet{BC03}. 
Meanwhile, \citet{Chang2015} obtained a stellar mass of $M_{s}$ = $10^{10.93}$M$_{\sun}$ 
for this galaxy by fitting the  spectral energy distribution to the combined SDSS and WISE photometry.
The optical radius of UGC~4056 is $R_{25}$ = 0.55 arcmin according to the
RC3 catalogue \citep{RC3}  or  $R_{25}$ = 0.49 arcmin according to the
HyperLeda\footnote{http://leda.univ-lyon1.fr/} database
\citep{Paturel2003,Makarov2014}.  
Furthermore, the optical radius of UGC~4056 is $R_{25}$ $\cong$ 21.26 kpc
or $R_{25}$ $\cong$18.95 kpc. We adopt here the mean value, $R_{25}$ = 20 kpc. 
The 2D spectroscopy of UGC~4056 was carried out within the MaNGA
survey project (its MaNGA datacube number is 8140-12703) \citep[its MaNGA datacube number is 8140-12703;]{Albareti2016}.

The publicly available data in MaNGA SDSS DR13 for the galaxy UGC~4056  
form the basis of the current study. The analysis was carried out as in
\citet{Zinchenko2016} and \citet{Pilyugin2018a,Pilyugin2018b}. Briefly,
for each spaxel spectrum, the fluxes of the emission lines were
measured. The velocity of each region (spaxel) is estimated from the
measured wavelength of the H$_\beta$ and H$_\alpha$ emission lines 
and from the stellar absorption lines.  The surface brightness in
the SDSS $g$ and $r$ bands was obtained from broadband SDSS images
created from the data cube.  
Those measurements were used for the determination of the photometric 
(surface brightness profile), the kinematic (inclination angle $i$,
the PA of the major axis, and the circular velocity curve), 
and the chemical (the abundance map and the radial abundance gradient) 
properties of UGC~4056.

\begin{figure*}
\resizebox{1.00\hsize}{!}{\includegraphics[angle=000]{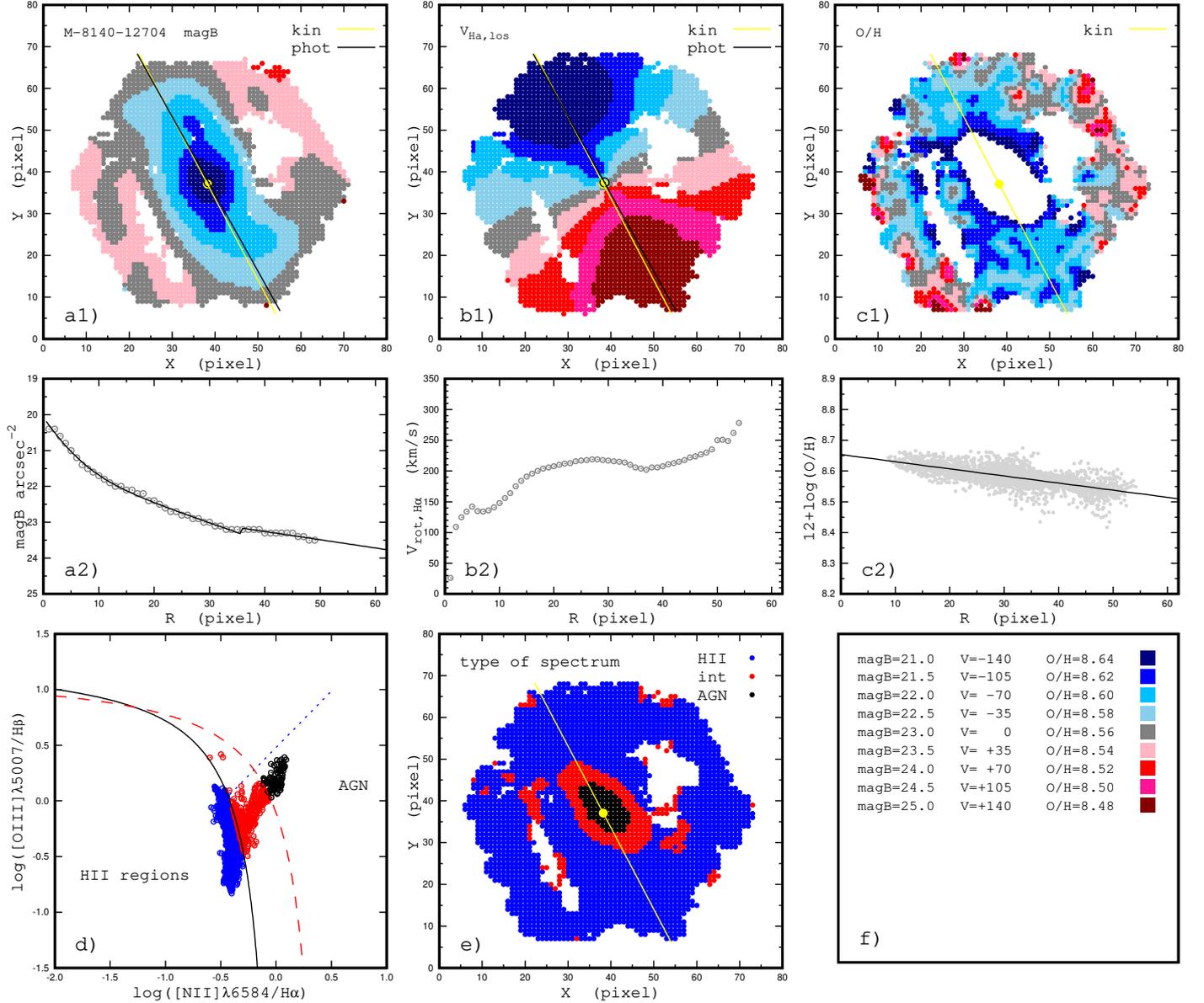}}
\caption{Distributions of the $B$-band 
surface brightness (panel $a1$), the observed (line-of-sight) 
H$_\alpha$ velocity (panel $b1$), and the oxygen abundance (panel $c1$) 
across the image of the galaxy UGC~4056 in sky coordinates (pixels). 
Each characteristic is colour-coded as explained in panel $f$. 
The circles show the positions of the kinematic
(yellow) and photometric (black) centre of the galaxy.
The lines indicate the positions of the corresponding major axis of the galaxy. 
Panel $a2$ shows the photometric profile constructed with kinematic 
angles (the inclination and the position of major axis; circles) 
and bulge-disc decomposition (line).
Panels $b2$ and $c2$ show the rotation velocity and the oxygen abundance as a function of radius, respectively.
Panel $d$ shows the BPT diagram for individual spaxels.
The solid (dark) and dashed (red) curves
mark the demarcation line between AGNs and H\,{\sc ii} regions defined
by \citet{Kauffmann2003} and \citet{Kewley2001}, respectively.  The
dotted (blue) line is the dividing line between Seyfert galaxies and
LINERs defined by \citet{CidFernandes2010}.  
Panel $e$ shows the distribution of spaxels with different types of radiation
in the spaxel spectra (H\,{\sc ii} region-like -- blue points,
intermediate -- red points, and AGN-like -- black points) over the image of the galaxy. 
Panel $f$ explains the colour-coding of the values of the $B$-band 
surface brightness (in units of mag arcsec$^{-2}$), the line-of-sight H$_\alpha$ velocity (in units of km s$^{-1}$), 
and the oxygen abundance (in units of dex) used in panels $a1, b1, c1$.
The pixel scale is 0.5 arcsec, the same as the size of the spaxels in the datacube.
}
\label{figure:maps}
\end{figure*}

\section{General properties of UGC~4056}
\label{section:properties}

The distributions of the surface brightness in the $B$ band, 
the observed (line-of-sight) H$_\alpha$ velocity, and the oxygen abundance 
across the image of UGC~4056 are presented in Fig.~\ref{figure:maps}.
The surface brightness in the $B$ band was obtained from the surface brightness 
in the SDSS g and r bands following the prescription described in \citet{Pilyugin2018a}.

Panel $a1$ of Fig.~\ref{figure:maps} shows the surface brightness distribution 
in the $B$ band across the image of UGC~4056 in sky coordinates (pixels).
The geometric parameters of UGC~4056 (the coordinates of the centre
$X_{0}$ = 38.5 pixels and $Y_{0}$ = 37.5 pixels, the inclination angle
$i$ = 64.2$\degr$, and the position angle of the major axis $PA$ = 208.4$\degr$)
were obtained in our current study from the analysis of the surface brightness map in the way
described in \citet{Pilyugin2018a}. The formal error is 0.5 pixels for $X_{0}$, $Y_{0}$ 
and 1$\degr$ for the inclination and PA.
The black circle shows the photometric centre of the galaxy and the
black line indicates the position of the major photometric axis of the galaxy.
The yellow circle shows the kinematic centre of the galaxy and the yellow line
indicates the position of its kinematic major axis, derived here following the method described in \citet{Pilyugin2018a},
also see detailed description in Section~\ref{sect:Rotation curve of M-8140-12703}) of the current study.

Panel $b1$ of Fig.~\ref{figure:maps} shows the observed H$_\alpha$
velocity $V_{los,{\rm H}{\alpha}}$ distribution across the image
of the galaxy in sky coordinates (pixels). The value of
the velocity is colour-coded with a step size of 35 km/s; dark-blue corresponds to 
the minimum negative velocity, dark-red corresponds to the maximum positive velocity,
and grey corresponds to zero line-of-sight velocity
(see colour code explanation in panel $f$ of Fig.~\ref{figure:maps}). 
Despite the generally regular appearance of the velocity field, 
panel $b1$ shows an unusually high velocity feature in the inner part of the galaxy.
Panel $b2$ shows the rotation velocity as a function of radius.

Panel $c1$ shows the oxygen abundance distribution across the image
of UGC~4056 in sky coordinates (pixels).
Panel $c2$ shows the radial distribution of the oxygen abundance.
The abundances were estimated using the $R$ calibration from \citet{Pilyugin2016}.
Since the $R$ calibration can be applied only to H\,{\sc ii} regions,
we used the [N\,{\sc ii}]$\lambda$6584/H$_\alpha$ versus\ [O\,{\sc iii}]$\lambda$5007/H$_\beta$
diagnostic diagram of \citet[the BPT diagram]{Baldwin1981} to distinguish between 
H\,{\sc ii} region-like  and AGN-like spectra (panel $d$ of Fig.~\ref{figure:maps}).
The solid (dark) and dashed (red) curves in panel $d$ of Fig.~\ref{figure:maps} 
mark the demarcation line between AGNs and H\,{\sc ii} regions defined
by \citet{Kauffmann2003} and \citet{Kewley2001}, respectively.  The
dotted (blue) line is the dividing line between Seyfert galaxies and
a low-ionization nuclear emission-line regions (LINERs) defined by \citet{CidFernandes2010}.
If a point is located to the left of (below) the Kauffmann et al. demarcation line then
this spectrum is classified as H\,{\sc ii} region-like spectrum (blue points).
If the point is located right (above) of the Kewley et al. demarcation line then
this spectrum is classified as an AGN-like spectrum (black points).
If the point is located between the Kauffmann et al. and Kewley et al. demarcation lines then
this spectrum is classified as intermediate (red points). Panel $e$ shows the spatial distribution
of zones with different types of radiation.
The oxygen abundance was estimated only for the spaxels below the Kauffmann et al. demarcation line.

\begin{figure*}
\resizebox{1.00\hsize}{!}{\includegraphics[angle=000]{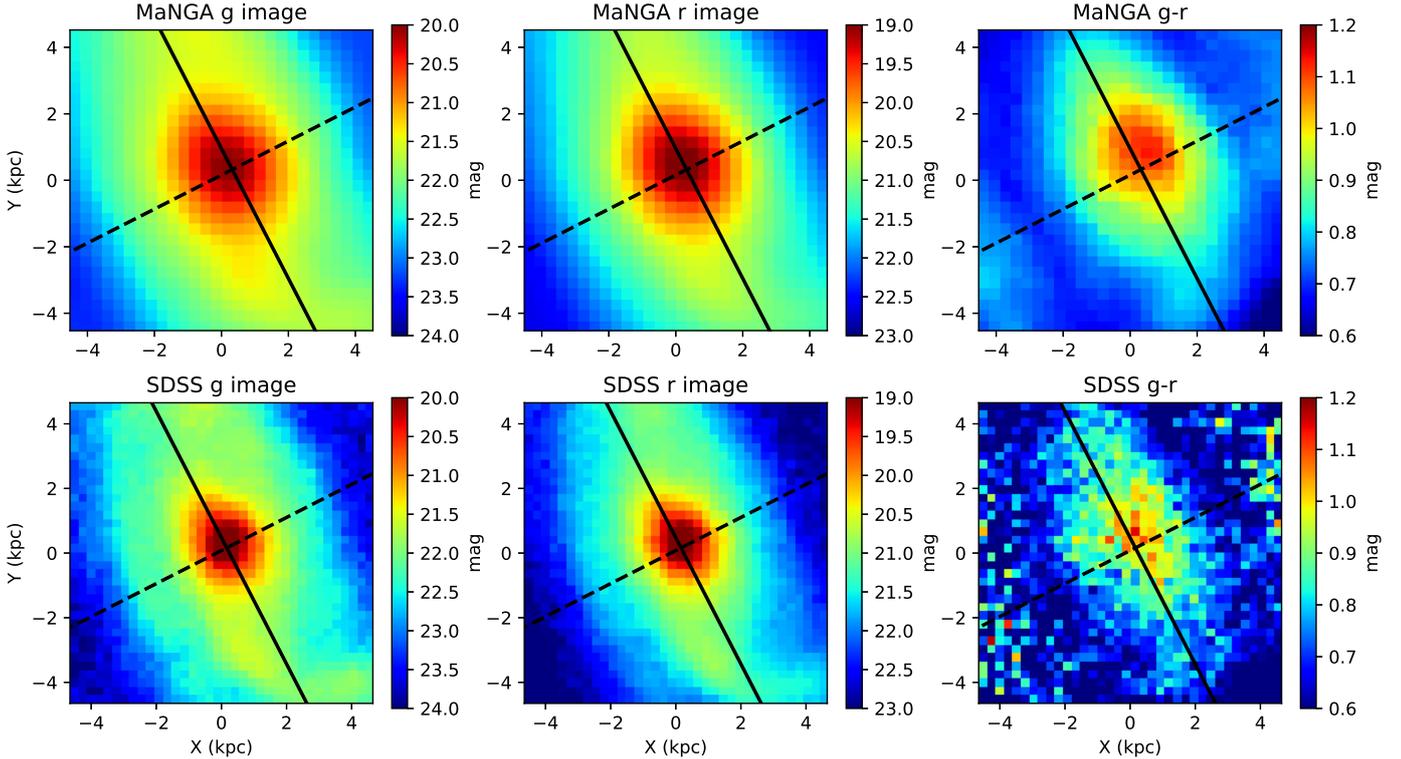}}
\caption{
Magnitudes in the g (left panels) and r (middle panels) bands and in the colour index g-r (right panels) 
from MaNGA and SDSS photometry in the inner part of UGC~4056. The solid and dashed lines indicate the 
major and minor axis of the galaxy, respectively.
}
\label{figure:g-r}
\end{figure*}

\begin{figure*}
\resizebox{1.00\hsize}{!}{\includegraphics[angle=000]{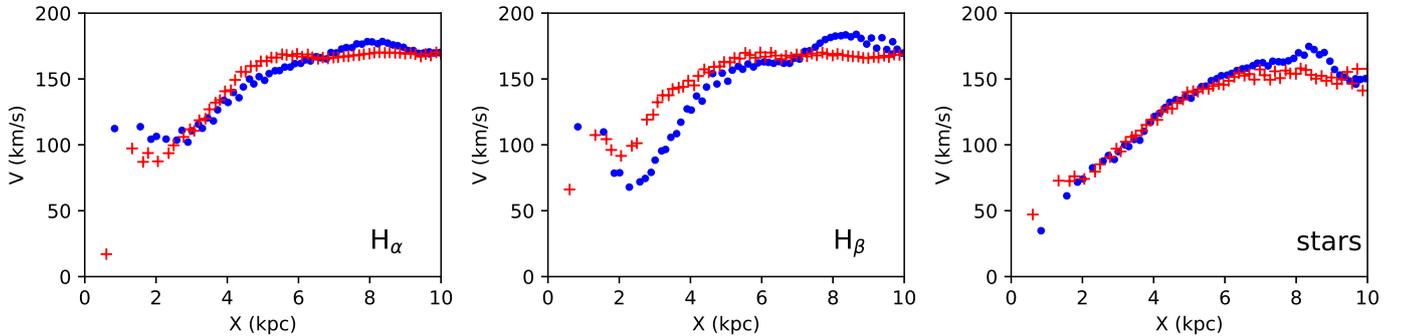}}
\caption{
Comparison of the approaching (blue circles) and the receding (red plus symbols) sides of the 
position--velocity diagram along the major axis.
}
\label{figure:Comparing_RC}
\end{figure*}

\section{The sense of rotation}
\label{sect:leading}

\citet{Thomasson1989} summarised three properties of an observed spiral galaxy 
that one needs to know in order to be able to decide whether its arms are leading or trailing:
(i) The direction of the spiral pattern (if it is S-or Z-shaped),
(ii) which side is approaching us, and 
(iii) which side is nearest to us.
Our galaxy is Z-shaped in Fig.~\ref{figure:image} with a blueshifted NE part and a redshifted SW part of the disc.
Thus, we need to determine the tilt of the galaxy, that is which side is closer to us.

It is possible to find the closer part of the galaxy from the analysis of the colour index map
of the galaxy assuming that the dust lane must be thinner than the stellar light distribution.
In this case, the closer side of a galaxy should have higher extinction and therefore a higher colour index.
Figure~\ref{figure:g-r} shows the maps of the SDSS and MaNGA colour index g-r (right panels) as well as
the magnitudes in the g and r SDSS bands (left and middle panels, respectively).
The solid and dashed lines are the major and minor axis of the galaxy, respectively.
As the maximum of the g-r colour index is clearly shifted to the right of the major 
axis, we conclude that the near side is to the right of the major axis.
Based on our knowledge of the approaching and receding sides, the rotation of the 
galaxy occurs clockwise on the sky plane and the spiral arms are trailing arms as in most galaxies.

\section{Morphology of peculiar motions in the central part of UGC~4056}
\label{sect:Morphology}

In this section, we analyse the diagram "position vs. velocity" along the minor 
axis and the major axis, which allows us to detect the peculiar radial and tangential motions in the 
galaxy without restrictions imposed by any kinematic model.
For this purpose, we used our measurements of the position angle, inclination, and
coordinates of the centre of the galaxy obtained from galaxy photometry in Section~\ref{section:properties}.

\begin{figure*}
\resizebox{1.00\hsize}{!}{\includegraphics[angle=000]{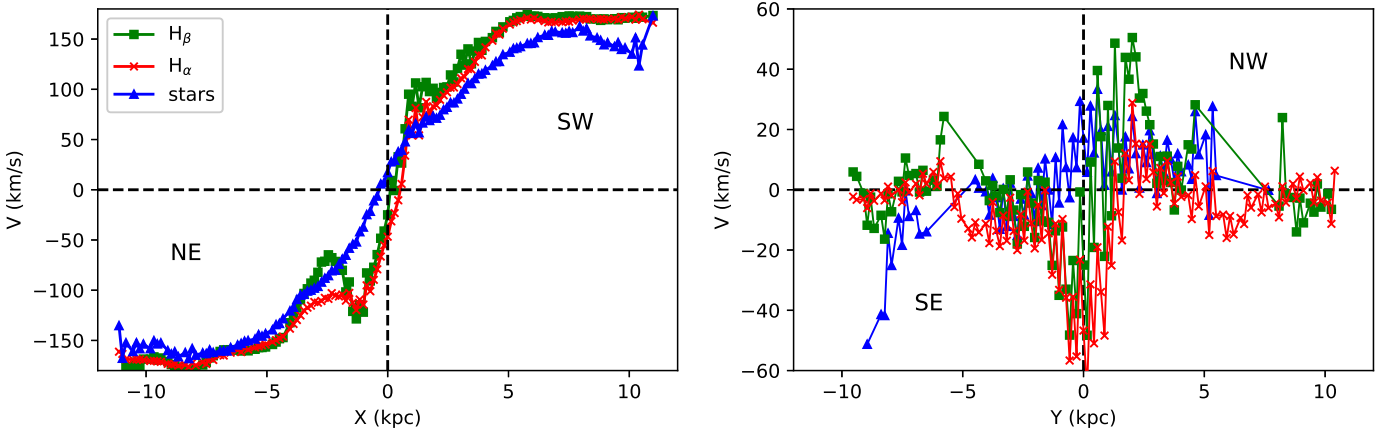}}
\caption{
Left panel: Distribution of the measured H$_\alpha$ (red crosses), H$_\beta$ (green squares), and 
stellar (blue triangles) line-of-sight velocities $V_{los}$ along the major axis over the  sky plane.
Right panel: Distribution of the measured H$_\alpha$ (red crosses), H$_\beta$ (green squares), and  stellar (blue triangles)
line-of-sight velocities $V_{los}$ along the minor axis over the  sky plane.
The measured points do not lie further than 0.32 kpc ($<$ 1 spaxel or  $<$ 0.5 arcsec) from the axes.
}
\label{figure:X_V_los-Ha_Hb}
\end{figure*}

\subsection{Line-of-sight velocities along the major axis}

The velocity of each region (spaxel) is estimated from the
measured wavelength of the H$_\beta$ and H$_\alpha$ emission line 
and from the stellar absorption lines, corrected for the redshift $z=$ 0.032104, 
which corresponds to a radial velocity of $V=$ 9625 km/s.
If the correction for the systemic radial velocity is right, 
then the measured line-of-sight velocities along the approaching side 
of the major axis must coincide with the ones along the receding side.
Figure~\ref{figure:Comparing_RC} compares the approaching (blue curve) 
with the receding (red curve) sides of the position--velocity diagram.
The curves of the approaching and the receding circular velocities generally coincide,
which confirms the estimated systemic radial velocity of the galaxy 
as well as the estimated coordinates of the centre of the galaxy.

We now consider  some features of the radial distribution of the measured velocities along the major axis.
In the nuclear region (R$<$2 kpc) the velocities of the ionized gas show a prominent `bump', 
while the velocities of stars increase more smoothly with radius. 
The high-velocity gas detected along the major axis in the central region can be explained as an impact
of the bar. The bisymmetric model of the bar by \cite{Spekkens2007} implies tangential and radial 
components of the non-circular motion caused by the bar, which can explain the streaming motions along both 
the major and minor axes. However, it should be noted that UGC 4056 has a weak bar and no such 
high-velocity gas streaming motions were found in the two barred galaxies from the sample of 
\citet{sakhibov2018}.

We  next consider the interval 2 kpc $<$R$<$4 kpc. Here the curves of the approaching and 
the receding velocities measured from the H$_\alpha$ line (Fig.~\ref{figure:Comparing_RC}, left panel)
 oscillate relative to each other along the major axis with an amplitude $<$ 10 km/s. 
Such oscillations can indicate peculiar velocities of gas with an amplitude $\approx$ 10 km/s 
caused by a symmetric second mode (two-armed spiral) of a spiral density wave in the galaxy disc.
On the other hand, in the case of measurements from the H$_\beta$ line (Fig.~\ref{figure:Comparing_RC}, middle panel) 
the receding velocity exceeds the approaching velocity in the entire interval from 2 kpc up to 4 kpc by $\approx$ 20 km/s. 
At galactocentric distances of R$>$4 kpc the curves of the approaching and the receding velocities
measured from the H$_\beta$ line (Fig.~\ref{figure:Comparing_RC}, middle panel) 
oscillate relative to each other in a similar fashion as in the case of the H$_\alpha$ line.
Curves of the approaching and receding velocities of stars ( Fig. ~ \ref {figure:Comparing_RC}, right panel) 
do not show noticeable deviations from each other.
The impact of spiral arms can be hidden because of the larger radial velocity dispersion of the stars.

Figure~\ref{figure:X_V_los-Ha_Hb} (left panel) shows  the distribution of the measured line-of-sight velocities $V_{los}$ 
of ionized gas (H$_\alpha$ -red crosses, H$_\beta$ -green squares) 
along the major axis in the approaching and receding sides of the galaxy.
We  note that the observed line-of-sight velocities are not corrected for the galaxy inclination. 
The gas `velocity bump' takes place in the central region ($R<$ 2 kpc) of the disc.
The distribution of the measured stellar line-of-sight velocities $V_{los}$ (blue triangles)
shows no peculiarity (no bump) in the central region ($R<$ 2 kpc) of the disc.
There is a systematic deviation between the circular velocity of the gas 
and circular velocity of stars in the area 4-10 kpc. 
The stellar disc rotates approximately 10-25 km/s more slowly than the gas disc. 
The slower rotation of the stars compared to the gas in the disc may be caused by 
the asymmetric drift \citep{Binney2008,Sysoliatina2018}.

\subsection{Line-of-sight velocities along the minor axes}

Because the points located on the minor axis of the galaxy do not contribute to the measured 
rotation velocity, but only to the radial component, 
it is interesting to examine the existence of any radial motion along the minor axis.
This statement is valid if the disc is relatively thin.
The line-of-sight velocities of stars (blue triangles) in the inner part of the minor axis (R$<$5 kpc) 
show a negligible radial component, which is approximately equal to the observational error.
In contrast, the gas shows a noticeable radial velocity in the inner (R$<$2 kpc) part of the minor axis (crosses and squares).
We plot only spaxels in which the flux signal-to-noise ratio is $\geq 6$.
We note that the observed velocities $V_{los}$ are not corrected for the galaxy inclination. 
A correction for the galaxy inclination increases the radial velocities up to $\approx 60$ km/s.
There are prominent radial velocities of stars at the south-east end of the minor axis.
The perturbation of the velocity field of stars at R$>$9 kpc is noticeable also when one compares
velocities of the approaching and receding sides of the disc (Fig.~\ref{figure:Comparing_RC}, right).

A negative radial velocity, measured in the distant part of the minor axis (SE part), 
and a positive radial velocity, measured in the closer part of the minor axis (NW part); see right panel of Fig.~\ref{figure:X_V_los-Ha_Hb}), 
indicates a significant inflow of ionized gas toward the central region ($R<3$ kpc) of the galaxy.

\begin{figure}
\resizebox{1.00\hsize}{!}{\includegraphics[angle=000]{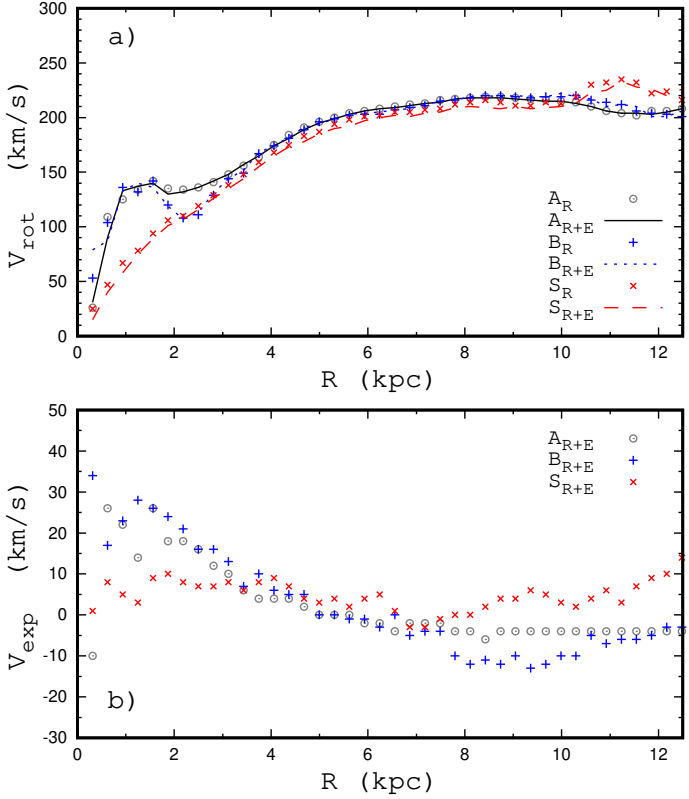}}
\caption{
  Circular velocity curves derived using all the spaxels with measured velocities (panel $a)$ .
  The circular velocity curve based on the H$_\alpha$ velocities is shown by the open circles
  (case $A_{R}$, the radial motion is neglected) and by the solid line 
  (case $A_{R+E}$, the radial motion is included). 
  The circular velocity curve based on the H$_\beta$ velocities is denoted by the plus symbols 
  (case $B_{R}$, without the radial motion) and by the dotted line 
  (case $B_{R+E}$, with the radial motion). 
  The circular velocity curve based on the stellar velocities is presented by the crosses 
  (case $S_{R}$, without the radial motion) and by the dashed line 
  (case $S_{R+E}$, with the radial motion). 
  Panel $b$ shows the radial velocity as a function of radius for
  the cases $A_{R+E}$ (open circles), $B_{R+E}$ (plus signs), and $S_{R+E}$ (crosses).
  The geometric parameters and other characteristics of each case are reported
  in Table \ref{table:rcs}.
}
\label{figure:rc-e}
\end{figure}

\begin{table*}
\caption[]{\label{table:rcs}
Conditions under which the different variants of the circular velocity curve are derived
(the tracer of the rotation velocity (2),
whether the radial motion is included or not (3),
the number of points used in deriving the rotation velocity (4)) and 
the obtained geometric parameters of UGC 4056 
(the position of the centre of the galaxy on the MaNGA image $X_{0}$ and $Y_{0}$ (5,6),
the position angle of the major axis $PA$ (7), 
the galaxy inclination angle $i$ (8), and 
the deviation $\sigma_{V_{los}}$ given by  Eq.~(\ref{equation:sigma})  (9)).
The number in parenthesis refers to the corresponding column.
}
\begin{center}
\begin{tabular}{lccccccccc} \hline \hline
notation                     &
tracer                      &
radial motion            &
N points                    &
$X_{0}$                       &
$Y_{0}$                       & 
PA                           &
$i$                          &
$\sigma_{V_{los}}$             \\
                            &
                             &
                             &
                               &                            
(pixel)                         &
(pixel)                         &
($\degr$)                      &
($\degr$)                      & 
(km/s)                      \\ 
(1)                            &
(2)                            &
(3)                            &
(4)                            &
(5)                            &
(6)                            &
(7)                            &
(8)                            &
(9)                           \\  \hline
A$_{R}$          &  $V_{{\rm H}{\alpha}}$  &  no                                                                             &  2818  &  38.2  &  37.1 &  207.1 &  52.9  &  6.36 \\ 
A$_{R+E}$        &  $V_{{\rm H}{\alpha}}$  &  included                                                                      &  2818  &  38.4  &  37.3 &  207.6 &  53.4   &  5.68 \\ 
                 &                     &                                                                                   &        &        &       &        &        &        \\ 
B$_{R}$          &  $V_{{\rm H}{\beta}}$  &  no                                                                           &  2818  & 38.1   & 37.9  &  206.4 &  54.2  & 11.26  \\ 
B$_{R+E}$        &  $V_{{\rm H}{\beta}}$   &  included                                                                        &  2818  & 37.9   & 37.9  &  207.9 &  53.9   & 10.32 \\ 
                 &                     &                                                                                   &        &        &       &        &         &       \\ 
S$_{R}$          &  $V_{star}$           &  no                                                                             &  2818  &  36.6  &  37.9 &  207.4 &  49.0   & 12.32 \\ 
S$_{R+E}$        &  $V_{star}$           &  included                                                                       &  2818  &  36.7  &  37.5 & 206.0 &  50.8   &  11.90\\ 
\hline
\end{tabular}\\
\end{center}
\begin{flushleft}
\end{flushleft}
\end{table*}

\section{Circular velocity curves of UGC~4056}
\label{sect:Rotation curve of M-8140-12703}

The main question is whether the velocity bump on the line-of-sight velocity curve 
(along the major axis in Fig.~\ref{figure:X_V_los-Ha_Hb}) 
displays a faster rotation of the ionized gas than the rotation of stars, 
or whether it is only a contribution to the line-of-sight line of the radial motion of the gas.
Below, we analyse 2D velocity maps of all our three tracers with models. 
The model considered in the present section considers simultaneously 
contributions of the tangential and radial components in the observed velocity (Eq.~\ref{equation:vxy}).
The model is able to separate the tangential and radial components of the observed velocity in every spaxel.
In Section~\ref{section:Fourier} we complement this model with terms 
that consider the impact of peculiar velocities caused by spiral arms.

\subsection{Determination of circular velocity curve}
\label{sect:Determination of rotation curve}

The determination of the rotation velocity from the observed velocity field
is performed in the standard way  \citep[e.g.][]{Warner1973,Begeman1989,deBlok2008,Oh2018}. 
The observed line-of-sight velocities $V_{los}$ recorded on a set of sky coordinates ($x,y$) are related 
to the kinematic  parameters by:
\begin{equation} 
V_{los}(x,y) = V_{sys} + V_{rot}\,\cos(\theta) \, \sin(i) + V_{rad}\,\sin(\theta) \, \sin(i)
\label{equation:vxy} 
.\end{equation} 
Here,
\begin{equation}
\cos(\theta)   =   \frac{-(x-x_{0})\, \sin(PA) + (y-y_{0}) \, \cos(PA)}{R} 
\label{equation:costeta}
,\end{equation}
\begin{equation}
\sin(\theta)   =   \frac{-(x-x_{0})\, \cos(PA) - (y-y_{0}) \, \sin(PA)}{R \cos(i)} 
\label{equation:sinteta}
,\end{equation}
where $R$ is the radius of a ring in the plane of the galaxy:
\begin{eqnarray}
       \begin{array}{lll}
 R & = & [ \{ - (x-x_0) \sin(PA) + (y-y_0) \cos(PA) \}^{2}  \\  
 & + &  \{[(x-x_0) \cos(PA) + (y-y_0) \sin(PA)]/\cos i\} ^{2} ]^{1/2}
\end{array}
\label{equation:rdisc}
,\end{eqnarray}
where $x_{0}$ and $y_{0}$ are the sky coordinates of the rotation centre of the galaxy, 
$V_{sys}$ is the systemic velocity, 
$V_{rot}$ is the circular velocity at the distance $R$ from the centre, 
$V_{rad}$ is the radial velocity at the distance $R$ from the centre, 
$PA$ is the position angle of the major axis, and 
$i$ is the inclination angle.

The deprojected galaxy plane is divided into rings of a width of one pixel.
The rotation velocity and the radial velocity are assumed to be the same 
for all the points within the ring.
The position angle of the major axis and galaxy inclination angle are assumed to be
the same for all the rings, that is, constant within the disc.
The parameters $x_{0}$, $y_{0}$, $V_{sys}$, $PA$, $i$, and the circular velocity curve $V_{rot}(R)$ were
derived through the best fit to the observed velocity field $V_{los}(x,y)$, that is, we require that
the mean deviation $\sigma_{V_{los}}$ given by
\begin{equation}
\sigma_{V_{los}} = \sqrt{ [\sum\limits_{j=1}^n (V_{los,j}^{cal} - V_{los,j}^{obs})^{2}]/n}  
\label{equation:sigma}
\end{equation}
is minimized. Here, the $V_{los,j}^{obs}$ is the measured line-of-sight velocity of the $j$-th spaxel and 
$V_{los,j}^{cal}$ is the velocity computed through Eq.~(\ref{equation:vxy}) for the corresponding
sky coordinates $x$ and $y$. 

\subsection{Circular velocity curve of UGC~4056}

\begin{figure*}
\resizebox{1.00\hsize}{!}{\includegraphics[angle=000]{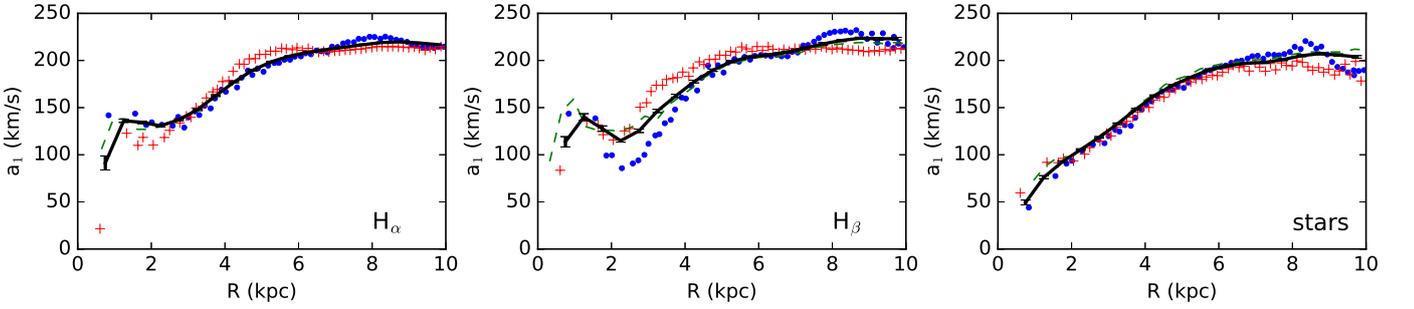}}
\caption{
The radial change of the Fourier coefficients $a_1$ (black line) computed from the H$_{\alpha}$,
H$_{\beta}$, and stellar component velocity maps. The green curve shows the rotation velocity computed 
in Section~\ref{sect:Rotation curve of M-8140-12703}. Blue circles and red plus signs show the 
rotation velocity along the approaching and the receding sides of the major axis 
corrected for the inclination of the galaxy.
}
\label{figure:F-Coeff_a1}
\end{figure*}

\begin{figure*}
\resizebox{1.00\hsize}{!}{\includegraphics[angle=000]{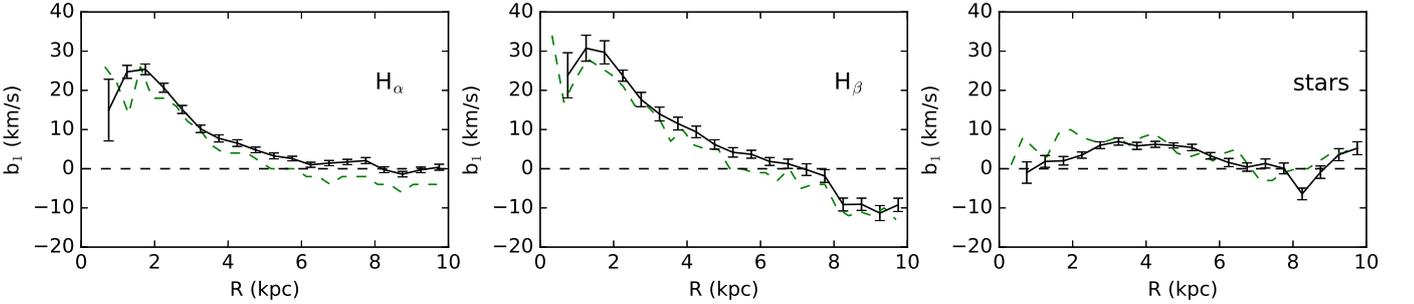}}
\caption{
The radial change of the Fourier coefficients $b_1$ (black line) computed from the H$_{\alpha}$,
H$_{\beta}$, and stellar component velocity maps. The green dashed curve shows the radial velocity computed 
in Section~\ref{sect:Rotation curve of M-8140-12703}. The black dashed line marks no radial motion.
}
\label{figure:F-Coeff_b1}
\end{figure*}

Three values of the line-of-sight velocity of each region (spaxel) are estimated
from the measured wavelengths of the H$_\alpha$ and H$_\beta$ emission lines  
and from the stellar spectra. 
The values of the H$_\alpha$, H$_\beta$, and stellar line-of-sight velocities were measured
in the spectra of around 2800 regions (spaxels) on the image of UGC~4056. 

Firstly, we derive the circular velocity curves using all the observed 
velocity values $V_{los}$ and neglecting a possible radial motion,
that is, we assume that the term  $V_{rad}$ in Eq.~(\ref{equation:vxy}) is 
equal to zero at any galactocentric distance. The circular velocity 
curves based on the observed H$_{\alpha}$ velocities $V_{los,{\rm H}{\alpha}}$ 
, the observed H$_{\beta}$ velocities $V_{los,{\rm H}{\beta}}$ 
, and the observed star velocities $V_{los,star}$ 
are referred to as cases  $A_{R}$ ,  $B_{R}$ , and $S_{R}$ , respectively. 
The geometric parameters of UGC~4056 obtained in those cases are reported
in Table \ref{table:rcs}. The derived circular velocity curves are shown in panel $a$ of Fig.~\ref{figure:rc-e} by open circles (case $A_{R}$), plus symbols
(case $B_{R}$), and crosses (case $S_{R}$). 

Examination of panel $a$ of Fig.~\ref{figure:rc-e} shows that the
$A_{R}$ circular velocity curve shows a hump at galactocentric distances 
between $R$ $\sim$ 3 and $R$ $\sim$ 10 pixels. 
The $B_{R}$ circular velocity curve is close to the $A_{R}$ curve and 
also shows a hump in this interval of galactocentric distances. 
In contrast to this, the rotation velocity of the stellar disc 
(the $S_{R}$ circular velocity curve) falls more or less smooth to zero with 
decreasing galactocentric distance in this interval. 
This suggests that the gas motion differs from the star motion 
in the central region of the galaxy UGC~4056.

It is evident that the uncertainties in the measured 
$V_{los,{\rm H}{\alpha}}$, $V_{los,{\rm H}{\beta}}$, and $V_{los,star}$
velocities could distort the corresponding circular velocity curves.
It was noted in \citet{Pilyugin2018b} that the velocity map obtained from the H$\alpha$ emission 
line measurements is preferable to derive the circular velocity curve of the gaseous disc. 
Indeed, the stronger line H$_\alpha$ is usually measured with higher precision than the weaker
H$_\beta$ line. One can therefore expect that the error in the measured wavelength 
of the H$_\alpha$ emission line $e(\lambda_{0})$ is lower than that for 
the H$_\beta$ emission line. Moreover, the error in the measured 
wavelength of the H$_\alpha$ line results in a lower error of the velocity
$e(\lambda_{0})$/$\lambda_{0}$ than the similar error in the measured 
wavelength of the H$_\beta$ line.  
Further, the error in the rotation velocity of the 
point (spaxel) $e(V_{rot})$ depends not only on the error in the line-of-sight
velocity $e(V_{los})$ but also on the position of the region 
(spaxel) in the galaxy (angle $\theta$). Those errors are related by the
expression  $e(V_{rot})$ = $e(V_{los})$/(cos$\theta$sin$i$) -- see Eq.~(\ref{equation:vxy}) --,  
that is, for a given error in the $V_{los}$ the error in the $V_{rot}$ is
minimum in the spaxels near the major axis of the galaxy and is
maximum in the spaxels along the minor axis. 
The MaNGA integral field unit covers only part of the galaxy UGC~4056.

As the next step, we derive the circular velocity curves taking into account 
the possible radial motion. 
The circular velocity curves based on $V_{los,{\rm H}{\alpha}}$ 
 , $V_{los,{\rm H}{\beta}}$ , and $V_{los,star}$  are referred to here as cases $A_{R+E}$ ,  $B_{R+E}$ ,
and  $S_{R+E}$ .
The derived circular velocity curves are shown in
panel $a$ of Fig.~\ref{figure:rc-e} by a solid line (case $A_{R+E}$), by a dotted line 
(case $B_{R+E}$), and by a dashed line (case $S_{R+E}$). 
The geometric parameters of the galaxy obtained in those cases are reported
in Table \ref{table:rcs}.
Inspection of panels $a$ and $c$ of Fig.~\ref{figure:rc-e} reveals 
a good agreement between the rotation velocities obtained with and without 
taking into account the radial motion.

Panel $b$ of Fig.~\ref{figure:rc-e} shows the radial velocity as a
function of radius for the cases $A_{R+E}$ (filled circles),  $B_{R+E}$ (plus signs),
and $S_{R+E}$ (crosses).
The disagreement between the values of the radial velocity at a given 
galactocentric distance obtained for the H$_\alpha$ and H$_\beta$ velocities 
can be as large as $\sim$10 km/s. This value can be adopted as the uncertainty 
in the $V_{rad}$ determinations. 
Panels $b$ and $d$ of Fig.~\ref{figure:rc-e} show that the 
gas radial velocities $V_{rad}$ at the central region can reach 30-40 km/s, 
that is they exceed the adopted uncertainty in the $V_{rad}$ determinations. 
The radial velocities of the stellar disc do not exceed the adopted 
uncertainty in the $V_{rad}$ at the central region. 
Thus, there is evidence in favour of the radial motion of the gaseous
disc near the centre.

\section{Fourier analysis of the azimuthal distribution of the non-circular motions in thin ring zones}
\label{section:Fourier}

The model used in Section~\ref{sect:Rotation curve of M-8140-12703} detects circular and axisymmetric radial motions. 
Moreover, the model provides the orientation parameters and the kinematical centre of the galaxy disc presented in Table~ \ref{table:rcs}.
In the current section, we investigate the impact of peculiar velocities caused by spiral density waves using Fourier analysis 
for the azimuthal distribution of the measured line-of-sight velocities in thin ring zones for a set of galactocentric distances.

\subsection{Model}
\label{section:Model}

As we discussed in our previous paper \citep{sakhibov2018},
the line-of-sight velocity $V^{obs}_r(R,\theta)$ observed at a
given point of the disc involves the velocity of the galaxy as a whole,
$V_{gal}$, the rotation velocity $V_{rot}(R)$, a pure radial
inflow or outflow $V_R$, and the peculiar velocities  caused by the density wave perturbation. 

We approximate the measured line-of-sight velocity $V^{obs}_r(R,\theta)$ 
at a given point ($R,\theta $) of the disc, using a similar approach as in \citet{sakhibov2018}:

\begin{eqnarray}
\label{equation:V_line-of-sight_2}
\frac{ V^{obs}_r(R,\theta)}{\sin i} = a_0+a_1 \cos(\theta)+ b_1 \sin (\theta)+\nonumber \\
+ a_2 \cos (2\theta)+ b_2 \sin (2\theta) + a_3 \cos (3\theta)+ b_3 \sin (3\theta) 
,\end{eqnarray}
where the coefficients $a_0,a_1,b_1,a_2,b_2, a_3$, and $b_3$ are 

\begin{eqnarray}
\label{equation:Fourieir_Coef}
a_0=V_{gal} +\frac{1}{2}(\hat u_1- \hat v_1 )\cdot \sin\Bigl(\cot \mu_1 \ln{(R/R_{01}) - \mu_1\Bigr)} \nonumber\\
a_1=V_{rot}(R)+ \frac{1}{2}(\hat u_2 - \hat v_2) \cdot \sin\Bigl(2\cot \mu_2 \ln{(R/R_{02}) - \mu_2\Bigr)} \nonumber \\
b_1=V_R - \frac{1}{2}(\hat u_2- \hat v_2) \cdot \cos\Bigl(2\cot \mu_2 \ln{(R/R_{02}) - \mu_2\Bigr)} \\
.
a_2= -\frac{1}{2}(\hat u_1+\hat v_1)\cdot \sin\Bigl(\cot \mu_1 \ln{(R/R_{01}) + \mu_1 \Bigr)} \nonumber \\
b_2= \frac{1}{2} (\hat u_1+\hat v_1) \cdot \cos\Bigl(\cot \mu_1 \ln{(R/R_{01}) + \mu_1 \Bigr)} \nonumber\\
a_3=-\frac{1}{2}(\hat u_2+\hat v_2)\cdot \sin\Bigl(2\cot \mu_2 \ln{(R/R_{02}) + \mu_2 \Bigr)} \nonumber \\
b_3= \frac{1}{2}(\hat u_2+\hat v_2) \cdot \cos\Bigl(2\cot \mu_2 \ln{(R/R_{02}) + \mu_2 \Bigr).} \nonumber
\end{eqnarray}
Here, $R$ and $\theta$ are polar coordinates of a point on the plane of the disc, 
the quantities $\hat u_1$, $\hat v_1$, $\mu_1$, and $R_{01}$
refer to the velocity perturbation amplitudes, the pitch angle, and
scaling factor for the first mode ($m=1$), and $\hat u_2$, $\hat v_2$,
$\mu_2$, and $R_{02}$ stand for the velocity perturbation amplitudes,
the pitch angle, and the scaling factor for the second mode ($m=2$) of
a spiral density wave, respectively.

Equation~(\ref{equation:V_line-of-sight_2}) and Eq.~(\ref{equation:Fourieir_Coef}) describe a model 
that takes into account not only the main rotation of the disc and radial symmetric motion,
but also the impact of the first and second modes of the spiral density wave in the disc.
A more detailed description and references were presented in \citet{sakhibov2018}.

\begin{figure}
\resizebox{1.00\hsize}{!}{\includegraphics[angle=000]{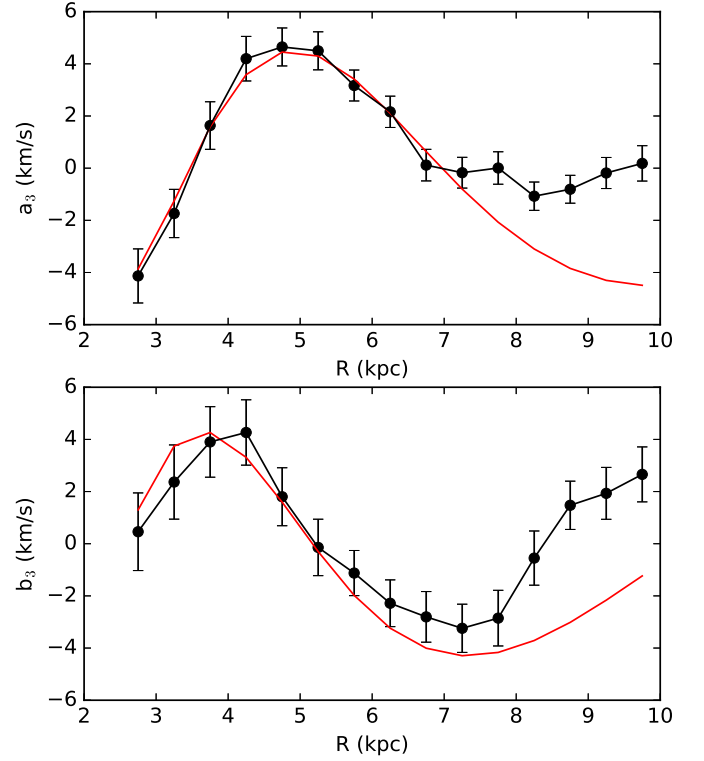}}
\caption{
Radial change of the third harmonics $a_3$ and $b_3$ compared with a two-armed spiral pattern 
with pitch angle $\mu_2 = 24^{\circ}$ and $R_{02} = 4$ kpc (red line). Filled circles show the 
derived values of the third harmonic. The red curves are a model for a two-armed spiral.  
}
\label{figure:Sp_2}
\end{figure}

\begin{figure}
\resizebox{1.00\hsize}{!}{\includegraphics[angle=000]{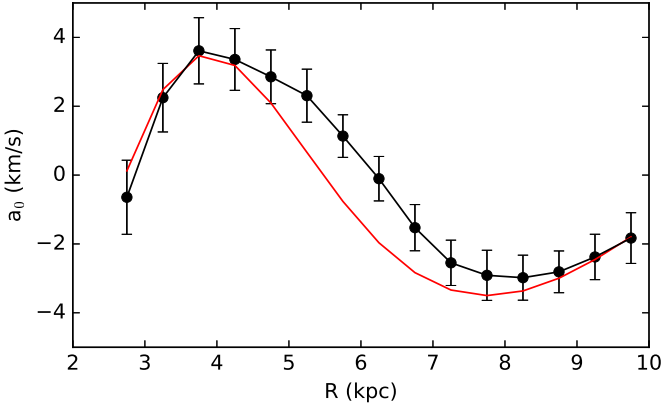}}
\caption{
Radial change of the zero harmonic $a_0$ compared with a model of the first mode 
of a spiral density wave with the pitch angle $\mu_1 = 12.6\degr$ and $R_{01} = 2.6$ kpc (red line). 
}
\label{figure:a0}
\end{figure}

\subsection{ Radial course of the Fourier harmonicas containing rotation and the radial motion}
\label{section:Radial course}

We carried out a Fourier analysis of the velocity maps derived from the measured wavelength 
of the H$_\alpha$ emission line and from the stellar absorption lines.

The black line in Fig.~\ref{figure:F-Coeff_a1} shows a radial course of the Fourier coefficients 
$a_1(R)$ computed from the H$_{\alpha}$, H$_{\beta}$, and stellar velocity maps.
As shown in  Eq.~\ref{equation:Fourieir_Coef}, the first Fourier coefficient $a_1$
includes contributions from the circular velocity $V_{\theta}$ and 
from the second mode of the spiral density wave. 
Since the amplitude of the velocity perturbation from the spiral arms is about 5$-$10 km/s 
and has the same magnitude as the accuracy of the circular velocity curve determined 
in Section~\ref{sect:Rotation curve of M-8140-12703} these two curves must coincide.
We compare the radial trend of the Fourier coefficient $a_1$  
with the rotation velocity computed in Section~\ref{sect:Rotation curve of M-8140-12703} 
(Fig.~\ref{figure:rc-e}, panel c) from Eq.~\ref{equation:vxy} (green line).
In general, the black and green curves coincide everywhere,
except for the central area of the disc (for gas) and the external border of the measured area of the disc.
The deviations between the two curves in these areas however are within the error interval.
Moreover, both curves show a `bump' at R$\approx$1 kpc in the case of ionized gas and a smooth course in the case of stars.
The blue stars and red crosses show the observations, the blue stars show approaching line-of-sight velocities observed along the major axis 
and corrected for inclination, and the red crosses show the receding ones.

Figure~\ref{figure:F-Coeff_b1} shows a radial course of the Fourier coefficients 
$b_1(R)$ (black line) computed from the H$_{\alpha}$, H$_{\beta}$, and stellar velocity maps.
We compare the radial change of the Fourier coefficient $b_1$  with the curve of radial velocity
computed in Section~\ref{sect:Rotation curve of M-8140-12703} 
(Fig.~\ref{figure:rc-e}, panel d) from Eq.~\ref{equation:vxy} (green line).
According to Eq.~\ref{equation:Fourieir_Coef} the coefficient $b_1$ contains the radially symmetric velocity $V_{R}$ 
and perturbations  from the spiral arms.
There is a difference between radial velocities of gas and stars in the central region of the disc (R$<$ 3 kpc). 
One can see that the ionized gas shows prominent radial motion up to $\approx$ 25 km/s at a 
galactocentric distance of R$\approx$ 2 kpc, while the stars show radial velocities $V_R<$ 10 km/s.
This corresponds to the absence of a bump on the circular velocity curve of stars 
and relatively small line-of-sight velocities of stars observed in the central part of 
the minor axis (Fig.~\ref{figure:X_V_los-Ha_Hb}, right panel, blue line).

Such gas motions at R $\approx$ 2 -- 4 kpc can be explained by the gas response to the bar potential.
Inside the co-rotation radius of the bar the gas may be accelerated along the leading edges of the bar.  
However, in the region R$<$ 1 kpc the bar potential should be close to axisymmetric, which could lead 
to various scenarios of inflow ranging from axisymmetric radial inflow to local streams of gas formed in 
the outer region of the bar.
As the spatial resolution of our data for UGC~4056 is $\sim$~1~kpc,
observations with the higher spatial resolution are needed to study the kinematics of the nuclear region in detail.

\subsection{Perturbations of the velocity field from spiral arms}
\label{section:Perturbations}

An impact of the spiral arms can be inferred through the radial change 
of the Fourier coefficients $a_3$ and $b_3$ presented in Fig.~\ref{figure:Sp_2}. 
As shown in Eq.~\ref{equation:Fourieir_Coef}, the third Fourier harmonic (coefficients $a_3$ and $b_3$) 
includes a contribution only from the second mode of the spiral density wave.
One can see that the velocity perturbations of the second mode of the spiral density wave 
change from about -5 to +5 km/s, which corresponds 
to a small-amplitude perturbation of the galactic gravitational potential ($\approx$  5 $\% $).
Figure~\ref{figure:Sp_2} shows that the periodic change of the third harmonic 
can be well approximated in the interval 2.5 $-$ 7.5 kpc with a two-armed (m=2) spiral pattern 
with the following geometrical parameters: pitch angle 
$\mu_2\approx 24^{\circ}$ and $R_{02}\approx$ 4 kpc.
We adopt a logarithmic form for the spiral arms
\begin{equation}
R =R_0 e^{\tan (\mu) \cdot \theta}
,\end{equation}
where $R$ and $\theta$ are polar coordinates in the plane of the
galaxy and $\mu$ is a pitch angle of the spiral arm. 

As shown in \citet{sakhibov2018}, the geometrical parameters of two-armed spirals $\mu_2$ and $R_{02}$
can be found from the radial change of the Fourier coefficients $a_3$ and $b_3$, 
using a linear regression 
\begin{equation}
y = A + Bx 
\label{equation:lin_regression}
,\end{equation}
with the notations
\begin{eqnarray}
\label{equation:lin-regression_coeff}
x =\ln(R) \nonumber \\
y = \arcsin\Biggl(\frac{a_3}{\sqrt{a_3^2+b_3^2}}\Biggr) \nonumber \\
A = \mu_2 -2\cot(\mu_2) \cdot \ln(R_{02}) \nonumber \\
B=2\cot(\mu_2). \nonumber
\end{eqnarray}
The estimation of the coefficients of the regression
(\ref{equation:lin_regression}) provides the geometrical parameters of
two-armed spirals, $\mu_2=24^{\circ} \pm 2^{\circ}$ and $R_{02}= 4 \pm 0.5$ kpc.
The same approach applied to the Fourier coefficients
$a_2(R)$ and $b_2(R)$ provides the geometrical parameters of a
one-armed spiral (mode m = 1), $\mu_1=13^{\circ} \pm 1^{\circ}$ and $R_{01}= 3 \pm 0.5$ kpc.

As the systemic velocity of the galaxy was subtracted before the calculation of the Fourier coefficients,
the radial course of a zero harmonic $a_0(R)$ corresponds 
to a radial change of the velocity perturbations from the first mode of a density wave.
Figure~\ref{figure:a0} shows that the periodic change of the zero harmonic $a_0(R)$ 
can be well approximated with a one-armed (m=1) spiral pattern with the following geometrical parameters: 
pitch angle $\mu_1 = 12.6^{\circ}$ and $R_{01}=$ 2.6 kpc, which conform to the $\mu_1$ and $R_{01}$
obtained using the radial distribution of the $a_2$ and $b_2$ Fourier coefficients.

\section*{Summary}
\label{section:Summary}

Using data from the MaNGA survey (datacube 8140-12703), we study the velocity field 
of the gas and stellar component in the giant spiral galaxy UGC~4056 which hosts an 
AGN. Using maps of the colour index $g-r$ obtained using 
SDSS and MaNGA images we 
determined the tilt of the galaxy, which allows us to conclude that
the galaxy rotates clockwise with trailing spiral arms.

We find a radial motion of gas in the central part of UGC~4056.
As shown in Section~\ref{sect:Morphology}, the radial motion 
of the ionized gas in the central part of the galaxy can produce an impact on the observed 
line-of-sight velocity curve (Fig.~\ref{figure:Comparing_RC} and Fig.~\ref{figure:X_V_los-Ha_Hb}, left).
Such an observational effect can be detected as a `velocity bump' on the central part of the circular velocity curve.
We demonstrate that the radial velocities detected in the central part of the galaxy  
can be caused by the inflow of the gas towards the nucleus of the galaxy.
This gas inflow can be explained by the response of the gas to the bar potential in the inner
part of the galaxy. This scenario supports the hypothesis that a bar could transport 
gas towards the inner kiloparsec \citep{Crenshaw2003}.

Along with a radial motion of gas in the central part of the galaxy, there is a gravitational spiral density wave 
that creates a field of peculiar velocities with an amplitude of about 5 km/s 
and that can be described by a logarithmic two-armed spiral with a pitch angle $\mu\approx 24^{\circ}$.
The pitch angle of $\approx 24^{\circ}$ corresponds to the late morphological type of UGC~4056.

Observations with high spatial resolution are needed to identify the detailed structure
and the cause of the peculiar velocity field in the central part of UGC~4056.

\section*{Acknowledgements}

We are grateful to the referee for his/her constructive comments. \\
I.A.Z.\ thanks the Max Planck Institute for Astrophysics for funding through
its visitor's program. \\
L.S.P., E.K.G., and I.A.Z.\ acknowledge support within the framework
of Sonderforschungsbereich (SFB 881) on ``The Milky Way System''
(especially subproject A5), which is funded by the German Research
Foundation (DFG). \\ 
L.S.P.\ and I.A.Z.\ thank for the hospitality of the
Astronomisches Rechen-Institut at Heidelberg University, where part of
this investigation was carried out. \\
I.A.Z. acknowledges the support of the National Academy of Sciences of Ukraine 
by the grant 417Kt and the support of the Volkswagen Foundation 
under the Trilateral Partnerships grant No.\ 90411. \\
P.B. acknowledges the support of the Volkswagen Foundation under the
Trilateral Partnerships grant 90411 and the support by the National 
Astronomical Observatories of Chinese Academy of Science (NAOC/CAS) 
through the Silk Road Project, through the Thousand Talents 
(“Qianren”) program and the President’s International Fellowship 
for Visiting Scientists and the National Science Foundation of 
China under grant No. 11673032. 
This work benefited from support by the International Space
Science Institute, Bern, Switzerland, through its International 
Team program ref. no. 393 ``The Evolution of Rich Stellar 
Populations \& BH Binaries'' (2017-18).
P.B. acknowledges the special support by the NASU under the
Main Astronomical Observatory GRID/GPU computing cluster project. \\
This work was partly funded by the subsidy allocated to Kazan Federal 
University for the state assignment in the sphere of scientific 
activities (L.S.P.).  \\ 
J.M.V. acknowledges support from the State Agency for Research of the 
Spanish MCIU through the ``Center of Excellence Severo Ochoa'' award for 
the Instituto de Astrof\'{i}sica de Andaluc\'{i}a (SEV-2017-0709), and from 
grant AYA2016-79724-C4-4-P cofunded by FEDER. \\
We acknowledge the usage of the HyperLeda database (http://leda.univ-lyon1.fr). \\
Funding for the Sloan Digital Sky Survey IV has been provided by the
Alfred P. Sloan Foundation, the U.S. Department of Energy Office of Science,
and the Participating Institutions. SDSS-IV acknowledges
support and resources from the Center for High-Performance Computing at
the University of Utah. The SDSS web site is www.sdss.org. \\
SDSS-IV is managed by the Astrophysical Research Consortium for the 
Participating Institutions of the SDSS Collaboration including the 
Brazilian Participation Group, the Carnegie Institution for Science, 
Carnegie Mellon University, the Chilean Participation Group,
the French Participation Group, Harvard-Smithsonian Center for Astrophysics, 
Instituto de Astrof\'isica de Canarias, The Johns Hopkins University, 
Kavli Institute for the Physics and Mathematics of the Universe (IPMU) / 
University of Tokyo, Lawrence Berkeley National Laboratory, 
Leibniz Institut f\"ur Astrophysik Potsdam (AIP),  
Max-Planck-Institut f\"ur Astronomie (MPIA Heidelberg), 
Max-Planck-Institut f\"ur Astrophysik (MPA Garching), 
Max-Planck-Institut f\"ur Extraterrestrische Physik (MPE), 
National Astronomical Observatories of China, New Mexico State University, 
New York University, University of Notre Dame, 
Observat\'ario Nacional / MCTI, The Ohio State University, 
Pennsylvania State University, Shanghai Astronomical Observatory, 
United Kingdom Participation Group,
Universidad Nacional Aut\'onoma de M\'exico, University of Arizona, 
University of Colorado Boulder, University of Oxford, University of Portsmouth, 
University of Utah, University of Virginia, University of Washington, University of Wisconsin, 
Vanderbilt University, and Yale University.


\end{document}